\documentclass[12pt,showpacs,preprintnumbers,prd,amssymb]{revtex4}
\usepackage{latexsym}
\usepackage{amsmath}
\usepackage{amsfonts}
\usepackage{amsbsy}
\usepackage{graphics}
\usepackage{graphicx}

\usepackage{color}

\begin{document}
\title{Iterative approach to the characteristic time\\  for chemical reactions of type \\ $A + B
\longleftrightarrow C + D$ \\ 
Homage to Alberto Santoro}

\author{R.  Aldrovandi }

\affiliation{Instituto de F\'{\i}sica Te\'orica \\
S\~ao Paulo State University  (UNESP) \\  S\~ao Paulo SP   Brazil}

\begin{abstract}
The analytic solution for the kinetic description of
binary reactions can be seen as the continuum version of a basic
discrete iterate mapping.  This fact allows a clear definition of  the
reaction characteristic time which takes the backward effect into account. 

\end{abstract}

\pacs{}

\maketitle

\section{Introduction}

In most usual laboratory and/or astrophysical conditions chemical 
reactions of type 
\begin{equation}
A + B \longrightarrow C + D \label{eq:reaction}
\end{equation}
are well described by the kinetic approach~\cite{Bal97}. 
If $n_{A}$, $n_{C}$ are the concentrations of species $A$ and
$C$, and $n = n_{A} + n_{C}$, the problem is fixed
by the relative concentrations
\begin{equation}
X_{A} = \frac{n_{A}}{n} \ \ \ {\textnormal{and}}\ \  X_{C} = \frac{n_{C}}{n} = 1 - X_{A} 
\ \ .
\end{equation}
Suppose some method is given for picking particles of types $A$ and $C$ 
while sampling  the system. Concentrations
$X_{A}(t)$ and $X_{C}(t)$ are then the relative probabilities of getting particles of
the corresponding species at time $t$. 

The reaction rates are typically given by the inverse times of free-flight,
or velocity/(mean free path)  ratios: if $v_{E}$ is the average velocity of
type-$E$ particles,
\begin{eqnarray}
R(A \rightarrow C) &=&  n_{A} \ v_{B} \ \sigma_{_{AB\rightarrow CD}}   = n\, v_{B} 
\,
\sigma_{_{AB\rightarrow CD}}   \, X_{A} \label{reactionrate11}\\
R(C \rightarrow A)  &=& n_{C} \ v_{D} \  \sigma_{_{CD\rightarrow AB}}  = n\,
v_{D} 
\, \sigma_{_{CD\rightarrow AB}} \, X_{C} \ ,\label{reactionrate21}
\end{eqnarray}
 where $\sigma_{_{AB\rightarrow CD}}  $ and $\sigma_{_{CD\rightarrow AB}}$ are the
corresponding  reaction cross-sections. The kinetic picture
underlying such definitions is well known~\cite{Born62,ZN71}: in
Eq.(\ref{reactionrate11}), for example,
$v_{B}\,
\sigma_{_{AB\rightarrow CD}}  $ is the effective cylindric volume presented by
particle
$B$ to particle $A$ per second. Particle $B$ will consequently meet 
$ n_{A} \ v_{B} \ \sigma_{_{AB\rightarrow CD}}  $  particles $A$  per second.

The relative concentrations as functions of time are then described
by the master (or gain/loss) equations
\begin{eqnarray}
\nonumber {\textstyle{\frac{d\ }{dt}}}  X_{C} &=&   R(A \rightarrow C) X_{A}
- R(C
\rightarrow A)  X_{C}\, ;
\\ 
\nonumber{\textstyle{\frac{d\ }{dt}}}  X_{A} &=& R(C \rightarrow A) X_{C} - 
R(A
\rightarrow C)  X_{A} \ .
\end{eqnarray}
Variation in the abundance of species $A$ is the  abundance of species $C$
times the rate of $C$-to-$A$ transformation (which represents the gain)
minus the $A$ abundance times the rate of its disappearance (the loss). 

\section{The solution}
\label{sec:solution}
Let us introduce the notations $a = n\, v_{D} \, \sigma_{_{CD\rightarrow AB}}$ 
and $b = n\, v_{B} \, \sigma_{_{AB\rightarrow CD}} $. Situations are not unusual
 in which both $a$ and $b$ are very nearly constant
(see Section \ref{sec:comments}). 
In that case, it is possible  to obtain general analytical solutions for the above
 master equations. The problem reduces to solving the
differential equation
\begin{equation}
{\textstyle{\frac{d\ }{dt}}} X_C(t) 
=   -\, a  X^2_C + b X^2_A = -\, a  X^2_C + b (1 - X_C)^2 =  
b + (b - a)\,  X^2_C - 2\, b\,  X_C  \label{masternp31}
\end{equation}
with constant coefficients. By their very meanings, $a > 0$ and $b > 0$.
Evolution will cease when
${\textstyle{\frac{d\ }{dt}}} X_C(t) = 0$, which suggests two candidate
equilibrium values:
$X^{(equil)}_C = \frac{b\pm \sqrt{ab}}{b-a}\,\,$. Once equilibrium is
attained, the backward reaction is as important as the forward reaction and Eq.(\ref{eq:reaction}) is, of course, better written with a
two-sided arrow,
\begin{equation}
A + B \longleftrightarrow C + D \label{eq:reactioninequil} \,\, .
\end{equation}

\begin{figure}[h]
\begin{center}
{\includegraphics[height=4.0cm,width=7.5cm]{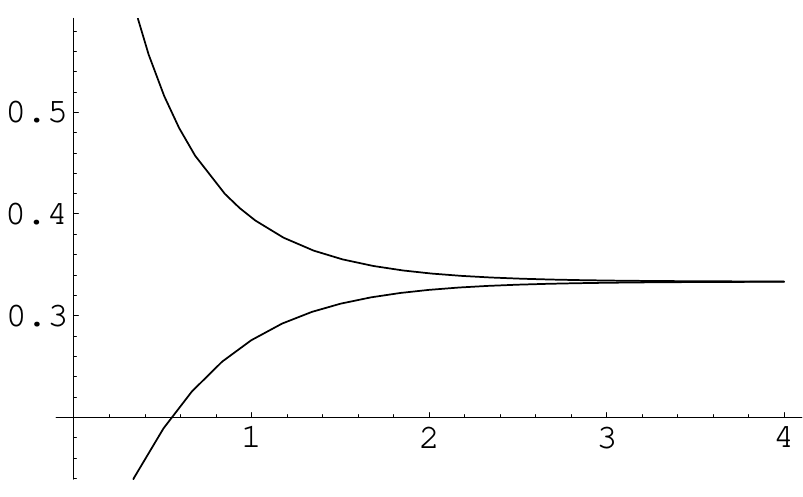}}
\caption{\em Relative C concentration, with  $a =  2$ and 
$b = 1/2$. Upper line: starting with no $A$; lower line: starting with
pure $A$. Equilibrium is
attained with $X_{C}(t \rightarrow \infty) = 1/3$. }
\label{ReactionAC}
\end{center}
\end{figure}

\begin{figure}[h]
\begin{center}
{\includegraphics[height=4.0cm,width=7.5cm]{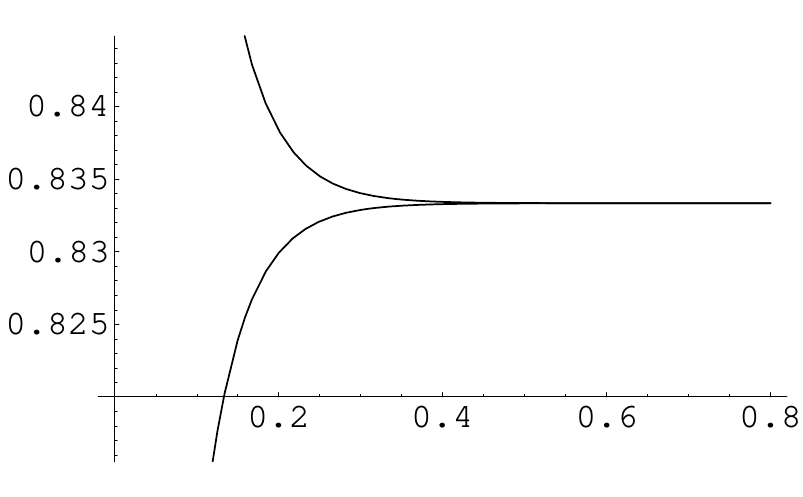}}
\caption{\em Same as the previous Figure, but with  $a =  2$ and 
$b = 50$.  Equilibrium is attained with $X_{C}(t \rightarrow \infty) = 5/6 = 0.8333 ... $. }
\label{ReactionAC2}
\end{center}
\end{figure}

The $X^2_C$ term in the right-hand side of (\ref{masternp31})  is present only if $a \ne b$. In that case the
solution is
\begin{equation}
X_{C}(t) = \frac{1}{b - a}\,\left\{b - {\sqrt{a\,b}}\,
     \tanh \left[{\sqrt{a\,b}}\,\, t + K \right]\right\} ,
\label{eq:solutionXC}
\end{equation}
with $K$ an integration constant whose determination will later provide our
main result. As lim$_{x\rightarrow
\infty}  
\tanh (x) = 1$, this solution tends indeed to one of the above candidate
equilibria:
\begin{equation}
X_C(\infty) = \frac{b - \sqrt{ab}}{b-a}
= \frac{\sqrt{b}}{\sqrt{b} + \sqrt{a}} = \frac{v_{B} 
\, \sigma_{_{AB\rightarrow CD}} - \sqrt{ v_{B} v_{D} \, \sigma_{CD\rightarrow
AB} \, \sigma_{_{AB\rightarrow CD}} }}{v_{B} 
\,\sigma_{_{AB\rightarrow CD}} \, -\,  v_{D} \, \sigma_{_{CD\rightarrow AB}}}\ \ .
\label{eq:equilibriumX} \end{equation}
The solution for $X_{A}(t) = 1 - X_{C}(t)$ is obtained by simply
exchanging parameters $a$ and $b$. Equilibria probabilities are related by
\begin{equation}
\frac{X_C(\infty)}{X_A(\infty)} = \frac{\sqrt{b}}{\sqrt{a}}  =
\frac{\sqrt{v_{B}\,\sigma_{_{AB\rightarrow CD}} }}{\sqrt{v_{D} \,
\sigma_{_{CD\rightarrow AB}}}} \, \, .
\label{eq:equilratio}
\end{equation}
Only to provide some intuitive guidance, plots for  toy models
$(a, b) =  (2, 1/2)$ and
$(a, b) =  (2, 50)$ are shown in Figures
\ref{ReactionAC} and \ref{ReactionAC2}.


An analogous result relates $X_B$ to $X_D$. Contact with the usual
equilibrium approach \cite{LL74,McQ76} can be made through a few
simple considerations.  In the equilibrium classical (nonrelativisitic, non--quantal) case,
particle $C$ (for example) will have concentration
$n_{C} = g_{C} \, \frac{e^{\mu_{C}/k T} }{\lambda_{C} ^{3}}$, where
$g_{C}$ counts the values taken by ``internal''  degrees of freedom (spin,
isospin, etc), $\lambda_{C}$ is the de Broglie thermal wavelength of
particle $C$ ($\lambda_C = \hbar \ \sqrt{\frac{2 \pi }{m_C k T}}$)
and $\mu_{C}$ is its chemical potential. The equilibrium
condition is $\mu_A + \mu_B = \mu_C + \mu_D$ or, in terms of the
fugacities $z = e^{\mu/kT}$,  $z_A z_B = z_C z_D$. Direct comparison with the above result
leads to
\begin{equation}
\frac{g_{C} g_{D}}{g_{A} g_{B}} 
\left[\frac{m_{C} m_{D}}
{m_{A} m_{B}}\right]^{3/2} = \sqrt{\frac{v_{A}
v_{B}}{v_{C} v_{D}}}\,\,\,
\frac{\sigma_{_{AB\rightarrow CD}}  }{\sigma_{_{CD\rightarrow AB}}} \,\, .
\end{equation}
If we use the equipartition formula  $v_B = \sqrt{\frac{3 k T}{m_B}}$, 
\begin{equation}
\frac{g_{C} g_{D}}{g_{A} g_{B}} 
\left[\frac{m_{C} m_{D}}
{m_{A} m_{B}}\right]^{5/4} = 
\frac{\sigma_{_{AB\rightarrow CD}}  }{\sigma_{_{CD\rightarrow AB}}} \,\, .
\end{equation}
We see that the cross-sections have, in this case, just to account for kinematic factors.
If we take na\"{\i}vely  $m_{A} = m_{C}$, $m_{B} = m_{D}$, $g_{A} =
g_{C}$, $g_{B} = g_{D}$,  the expected trivial equilibrium
requirement follows:  $\sigma_{_{AB\rightarrow CD}}   = \sigma_{CD\rightarrow
AB}$.

Situation $a = b$, which must be considered separately, would turn up in the
peculiar case
$v_{D} \, \sigma_{_{CD\rightarrow AB}}=  v_{B} 
\, \sigma_{_{AB\rightarrow CD}}  $:
the volume spanned by $D$ per unit time, as seen by $C$, equals the 
volume spanned by $B$ per unit time, as seen by $A$.
The solution of Eq.(\ref{masternp31}) would, in that case,  be
\begin{equation}
X_{C}(t) = \frac{1}{2} + e^{-\,2\,b\,t} (X_0 - \frac{1}{2})
\label{eq:specialcase}
\end{equation}
with, naturally enough, the
probabilities tending to equilibrium at $X_{C} = X_{A} = \frac{1}{2}$. An
example would be an ``elastic'' reaction of type
\begin{equation}
A + B \longrightarrow A + B \label{eq:reactioninequil} \ \ ,
\end{equation} 
with the same cross-section $\sigma$ in both sides. The crossed reaction $A + B
\longrightarrow  B + A$ would be accounted for by the general case, as $a =
n v_A \sigma \ne b = n v_B
\sigma$, equilibrium being given by the condition $\frac{X_B}{X_A} =
\sqrt{\frac{v_{B}}{v_{A}}}$ .

\section{Characteristic time}

There are two main approaches to evolving systems.  We have above used the
first: time evolution is described by a continuous curve of type
$X_t = f^{<t>}(X_0)$, solution of some  differential equation. In the second,
evolution is described by the successive iterations of a
mapping~\cite{Ras89,McC93,Ott94}.  The
state is known at each step, as if the ``time'' parameter of the system were
defined only at discrete values.  We can go from the first approach to the
second by taking the intersections leading to a Poincar\'e map. This
approach supposes a characteristic time --- the time of a unit step. If a
continuous description can be shown to be the interpolation of a discrete
mapping~\cite{Ald01}, a clear notion of  characteristic time obtains. There
is, however, a strong requirement: that interpolation must preserve the
notion of iteration all along. This  requirement is encapsulated in the
so-called semigroup conditions~\cite{AF98}. For a function $f(x, t) \equiv
f^{<t>}(x)$ describing the dynamical flow of a system, these conditions are
\begin{gather} 
f^{<t>}[f^{<t'>}(x)] = f^{<t'>}[f^{<t>}(x)] = f^{<t+t'>}(x) \; 
\; ; \label{I.1}
   \\
f^{<0>}(x) = x \; . \label{I.2}
\end{gather} 
A sufficient condition for that is that the solution have the form
\begin{equation} %
g(x, t) = F^{<-1>}\left[ c^{t} F(x)\right]  ,
\label{contiterateBriggs} 
\end{equation} %
for some function $F(x)$, its inverse $F^{<-1>}(x)$ and a constant $c$. This
would mean that $F(x)$ solves the  Schr\"{o}der  functional  equation
\begin{equation} 
F[g(x)]= c \; F(x).
\end{equation}
This can be translated into the additive form $f[g(x)] = c' + f(x)$ by taking
$f(x) = 
\ln [F(x)]$ --- what matters is that the  semigroup
conditions be respected.

Let us now notice that the integration constant $K$ in
 (\ref{eq:solutionXC}) can be obtained by simply taking the inverse function
at $t=0$. That solution assumes then the form
\begin{equation}
X_{C}(t) = \frac{1}{b - a}\,\left\{b - {\sqrt{a\,b}}\,
     \tanh \left[{\sqrt{a\,b}}\,\, t + 
     {\textnormal{arctanh}}\left(\frac{b + 
           \left(a - b\right)
\,{X_{C}(0)}}{{\sqrt{a\,b}}}\right)\right]\right\} . \label{eq:solutionXC2}
\end{equation}
This is actually the continuum form of
an iterate discrete mapping, and fulfills the  semigroup requirement. 
In more detail: introduce the notations
$f \circ g$ for the composition of functions $f$ and $g$, $f^{<m>}$ for the
$m$-th iterate of $f$ and
$f^{<-1>}$ for the function inverse to $f$. Then, with the functions
\begin{equation}
w^{<-1>} (z) = \frac{1}{a - b}\,\left(-b + {\sqrt{a\,b}}\,\tanh z \right) , 
\ v  = w^{<-1>} \circ f \circ w   ,\ f(u) = \sqrt{a\,b} + u,
\end{equation}
 expression (\ref{eq:solutionXC2})
is in effect the continuum version of  
\begin{equation}
X_{p}(m) = v^{<m>} \left(X_{p}(0) \right) = w^{<-1>} \circ f^{<m>}  \circ
w \left(X_{p}(0)
\right) .
\label{eq:iterateversion}
\end{equation}
The quantity 
\begin{gather} 
\tau = (a b)^{-1/2} = \frac{1}{n\, \sqrt{v_{B} v_{D} \,
\sigma_{_{CD\rightarrow AB}} \, \sigma_{AB\rightarrow
CD}} }\label{eq:chartime}
\end{gather} 
is the one-step time in the iteration and indicates the time interval in
which the reaction process does make significant progress.

Solution (\ref{eq:specialcase}) for the special  $a = b$ case has already
been written in iterative form, with $w^{<-1>} (z) = \frac{1}{2} + z$ and
characteristic time $\tau = (2 b)^{-1} = (2 n\, v_{B} 
\, \sigma_{_{AB\rightarrow CD}}  )^{-1} = (2 n\, v_{D} \, \sigma_{CD\rightarrow AB})^{-1} $, half the time of free flight.

\section{Final comments}
\label{sec:comments}
The assumption  used above~---~that $a$ and
$b$ are constant~---~actually mean  that  time
 $(a b)^{-1/2}$ is short in comparison with any other ``macroscopic''
time-scale involved. Such a  ``macroscopic'' time can, for instance, be the time
in which temperature and/or volume of the system change appreciably under the influence of some external agent. For reactions of cosmological interest, as those
involved in primordial nucleosynthesis, it is the inverse rate of expansion,
or inverse Hubble function. The quantities (velocities and cross sections)
appearing in  Eq.(\ref{eq:equilibriumX}) will depend on such large-scale time.
The first criterion for the validity of the above results is, consequently,
that   $(a b)^{-1/2}$ be very short in comparison to those times. In order to use  the equilibrium formulae, it would be necessary that 
equilibrium be attained in not too many steps. 

It is usual to take the average
time of free flight
$\tau = (n \sigma v)^{-1}$ as an order-of-magnitude indication  of the 
lapse necessary for thermalization to be established. It gives a 
rough measure of the time between two ``hits'' in the reaction. 
This parameter,  however,  turns up under conditions quite different from those supposed
above~\cite{Born62}. It is conceived for an arrangement by which particles
of type
$A$ impinge with constant velocity $v_A$ upon a medium formed by particles of
type $B$ with constant number density $n_B$. With (constant) cross section
$\sigma_{_{AB}}$, the mean free path of an $A$ particle in the medium will be
$\lambda_A$ = $\frac{1}{n_{_B} \sigma_{_{AB}}}$, and the corresponding time of
free flight, $\tau_A = \frac{\lambda_A}{v_A} = \frac{1}{n_B v_A
\sigma_{AB}}$. The number of unscattered (that is, keeping the
same momentum direction) particles $A$ will then be given by 
$$
\frac{d n_A}{dt} = -\,\, \frac{n_A}{\tau_A}\,\,\, .
$$
This is a pure-loss equation, quite the same as that for radiative decay. 
Eventual reproduction of particles $A$ with the original momentum by other
scatterings is neglected, so that there is no gain. The solution is, of
course, the radiative decay formula
$$
n_A(t) = n_A(0)\, e^{-\, \frac{t}{\tau_A}}\,\,\, .
$$

This corresponds, up to normalizations, to the special solution
(\ref{eq:specialcase}). Expression (\ref{eq:chartime}), 
$$
\tau = \frac{1}{(n_A + n_C) \sqrt{v_{_B} v_{_D} \sigma_{_{AB\rightarrow
CD}} \sigma_{_{CD\rightarrow AB}}}}  =   \frac{ \sqrt{n_A n_C} }{n\, \sqrt{R(A \rightarrow C)\, R(C \rightarrow A)} }\, ,
$$
coming from gain-loss  considerations, takes also the backward reaction into account and provides, in principle at least,  a far better measure of  the
reaction characteristic time.

\end{document}